\documentclass[a4paper,english]{elsarticle}
\usepackage{booktabs}
\usepackage{amsmath}
\usepackage{amssymb}
\usepackage{graphicx}
\usepackage{upgreek}
\usepackage{url}
\usepackage{lineno}
\usepackage{a4wide}
\usepackage{subcaption}

\begin{document}

\title{Study of point- and cluster-defects in radiation-damaged silicon}

\renewcommand{\thefootnote}{\fnsymbol{footnote}}

\author[]{Elena~M.~Donegani$^{1}$}
\author[]{Eckhart~Fretwurst$^{1}$}
\author[]{Erika~Garutti$^{1}$}
\author[]{Robert~Klanner$^{1}$ \corref{cor1}}
\author[]{Gunnar~Lindstroem$^{1}$}
\author[]{Ioana~Pintilie$^{2}$}
\author[]{Roxana~Radu$^{2}$}
\author[]{Joern~Schwandt$^{1}$}

\cortext[cor1]{Corresponding author. Email: Robert.Klanner@desy.de. Telephone: +49 40 8998 2558.}

\address{$^{1}$Institute for Experimental Physics, University of Hamburg, Luruper Chaussee 149, 22761 Hamburg, Germany\\$^{2}$National Institute of Material Physics, Magurele, Romania}

\begin{abstract}
 Non-ionising energy loss of radiation produces point defects and defect clusters in silicon, which result in a  significant degradation of sensor performance.
 In this contribution results from TSC (Thermally Stimulated Current) defect spectroscopy for silicon pad diodes irradiated by electrons to fluences of a few~$10^{14}$\,cm$^{-2}$ and energies between 3.5 and 27\,MeV for isochronal annealing between 80 and 280$^{\,\circ}$C, are presented.
 A method based on SRH (Shockley-Read-Hall) statistics is introduced, which assumes that the ionisation energy of the defects in a cluster depends on the fraction of occupied traps.
 The difference of ionisation energy of an isolated point defect and a fully occupied cluster, $\Delta E_a$, is extracted from the TSC data.

 For the VO$_i$ (vacancy-oxygen interstitial) defect $\Delta E_a = 0$ is found, which confirms that it is a point defect, and validates the method for point defects.
 For clusters made of deep acceptors the $\Delta E_a$\,values for different defects are determined after annealing at 80$^{\,\circ}$C as a function of electron energy, and for the irradiation with 15\,MeV electrons as a function of annealing temperature.
 For the irradiation with 3.5\,MeV electrons the value $\Delta E_a = 0$ is found, whereas for the electron energies of 6 to 27\,MeV $\Delta E_a > 0$.
 This agrees with the expected threshold of about 5\,MeV for cluster formation by electrons.
 The $\Delta E_a$\,values determined as a function of annealing temperature show that the annealing rate is different for different defects.
 A naive diffusion model is used to estimate the temperature dependencies of the diffusion of the defects in the clusters.

\end{abstract}

\maketitle
 \tableofcontents
  \newpage
\section{Introduction}
 \label{sect:introduction}
\begin{figure}[!b]
\centering
\includegraphics[width=0.5\linewidth]{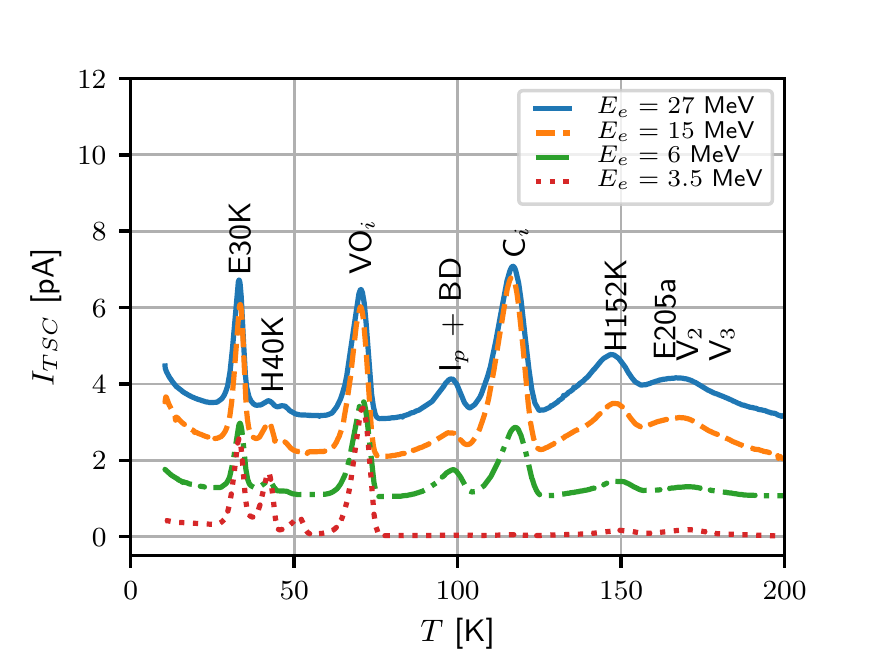}
 \caption{TSC spectra, $I_{TSC}(T)$, normalized to an electron fluence of $\Phi = 10^{14}$~cm$^{-2}$, after irradiation by electrons with different energies and annealing for 30~min at 80$^{\,\circ}$C, measured at a heating rate of 0.183\,K/s.
 The pad diodes were fabricated using standard float-zone silicon. For the trap filling forward-current injection has been applied at $T_0=10$~K, so that both electron~(E) and hole~(H) traps contribute to $I_{TSC}(T)$.
 For clarity, the individual curves are shifted vertically by 1~pA.
 For the identification and labeling of the individual defects we refer to~\cite{radu} and references therein. The dark current, which starts to dominate above 200~K, is subtracted.}\label{tsc_electrons}
\end{figure}

 Bulk-radiation damage in silicon limits the use of silicon detectors in high-radiation environments like at the CERN-LHC or in space.
 Although both microscopic and macroscopic effects of bulk damage are qualitatively understood, in spite of claims to the contrary, a consistent quantitative description of the data available has not yet been achieved.
 However, this is required for reliably predicting the sensor performance as a function of particle type and fluence, sensor design and operating parameters.
 The reason is that the number of radiation-induced states in the silicon band gap is large (see figure~\ref{tsc_electrons}), their properties are frequently only poorly known, and \emph{effective states}~\cite{eremin} have to be used in simulations because of the large number of defects.
 In addition, a quantitative understanding of defect clusters is lacking and in the TCAD simulations they are approximated by point defects.

 In this contribution we propose a simple, physics motivated parametrization of the properties of cluster defects and apply it to spectroscopic results from TSC (Thermally Stimulated Current) measurements.
 We present results for silicon irradiated with electrons of 3.5 to 27~MeV kinetic energy~\cite{radu,radu2}.
 As the threshold for defect cluster production is expected to be around 5\,MeV for electrons\,\cite{Lint}, these data are well suited to check the validity of the method.

 In the present work single vacancy related defects are considered to be point-like defects (e.\,g.\ the VO$_i$ defect).
 Such isolated point defects are produced in silicon by low energy recoils, whereas an agglomeration of defects (or a cluster for short) results from high energy recoils that introduce a dense cascade of silicon atoms displaced from their original lattice position.

\section{Test Structures and Measurements}
 \label{sect:measurements}
 For the studies p$^+$n~n$^+$ pad diodes, produced on 283~$\upmu$m thick float-zone n-type silicon with a phosphorous-doping of approximately 10$^{12}$~cm$^{-3}$ and an oxygen concentration $< 10^{16}$\,cm$^{-3}$, were used.
 The p$^+$ implant of 25~mm$^2$ area is surrounded by a guard ring.
 A window in the aluminum on top of the p$^+$\,implant with the shape of a decagon and an area of 3.25\,mm$^2$ allows to inject light through the p$^+$\,contact.
 The n$^+$\,back contact is covered by an aluminum grid.

 The pad diodes have been irradiated with electrons of $E_e = 3.5$, 6, 15, and  27\,MeV to fluences between 1 and $10 \times10^{14}$~cm$^{-2}$.
 For the sample irradiated with 15~MeV electrons, TSC measurements were performed before and after isochronal annealing for 30~min at temperatures $T_{ann}=80-280 ^{\,\circ}$C, in 20$^{\,\circ}$C steps.
 At the other energies, data were only taken before and after annealing of 30~min at $T_{ann}= 80 ^{\,\circ}$C.
 For the TSC measurements the bias was applied to the back n$^+$\,contact, and both p$^+$\,contact and guard ring were at ground potential.

 Figure~\ref{tsc_electrons} shows typical TSC spectra of pad diodes irradiated by electrons of different  energies after 30~min annealing at 80$^{\,\circ}$C.
 Trap filling has been applied at $T_0=10$~K with a forward-current of approximately 1~mA, so that both electron- and hole-traps are filled and contribute to the TSC~spectra.
 The heating rate was $\beta = 0.183$\,K/s.
 For more details we refer to~\cite{radu,junkes,pinti2002}.
 In the following we will limit the discussion to the point defect VO$_i$ (vacancy-oxygen interstitial) at 70~K and to the states in the $120 - 200$~K region, which are known to
 be cluster defects and have a significant impact on the sensor performance~\cite{ultimo}.

\begin{figure}
\centering
\includegraphics[width=0.5\linewidth]{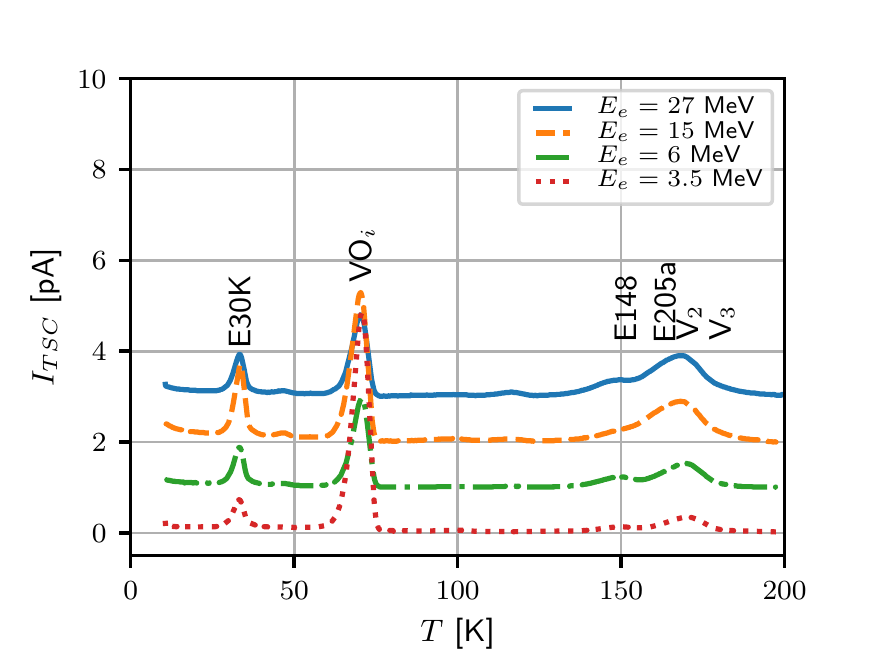}
 \caption{TSC spectra, $I_{TSC}(T)$, obtained after trap filling with light injection (with $\lambda =520$~nm) through the 3.25\,mm$^2$ window of the p$^{+}$ contact  at $T_0 = 10$~K, for the same irradiation conditions and samples as the data shown in figure~\ref{tsc_electrons}.
 As a result of the trap filling by light injection, only electron traps contribute to the TSC current. For clarity, the individual curves are shifted vertically by 1~pA.}\label{tsce}
\end{figure}

 For the TSC measurements analysed in this paper, the diodes were cooled to $T_0 = 10$~K at 200~V reverse bias to assure empty traps.
 At the temperature $T_0$ the traps were filled with electrons by injecting light of 520~nm through the 3.25\,mm$^2$ window of the p$^{+}$ contact, and the current, $I_{TSC}$, released by the traps was recorded as a function of $T$ for a constant heating rate $\beta = 0.183$~K/s.
 As a result of the light injection, only  electron traps are filled and only the charges released from electron traps contribute to $I_{TSC}$ (see figure~\ref{tsce}).

\section{Analysis Method}
\label{sect:method}
 According to SRH statistics~\cite{shockley,hall} the temperature dependence of the TSC current, $I_{TSC}(T)$, from acceptor states at an energy $E_a$ from the conduction band, which are filled with electrons at the temperature $T_0$ to the concentration $N_t$, is given by:
\begin{equation}
\label{srheq}
  I_{TSC}(T) = \frac{A \cdot d \cdot q_0}{2 } \cdot e(T) \cdot f_{t}(T) \cdot N_t,
\end{equation}
\begin{equation}\label{emission}
  e(T) = \sigma \cdot v_{th}(T) \cdot N_{C}(T) \cdot \textrm{exp} \left( - \frac{E_a}{k_B T} \right),
\end{equation}
\begin{equation}
 f_t(T) = \textrm{exp} \left( - \frac{1}{\beta} \int _{T_0} ^{T}  e(T^{'}) \, \mathrm{d}T^{'}\right),
\end{equation}
 with the elementary charge $q_0$, the diode volume where the traps have been filled $A \cdot d$, the emission rate $e(T)$, the ratio of filled states at the temperature $T$ relative to $T_0$, $f_t(T)$, the thermal velocity of electrons $v_{th}(T)$, the electron capture cross-section $\sigma$, and the density of states at the conduction band $N_C(T)$.
 For these quantities the default values of Synopsys TCAD~\cite{synopsys}, in particular the relation $v_{th}(T) = \sqrt{8 \cdot k_B T/\left(\pi \cdot m_{th}(300~\text{K})\right)}$
 of \cite{green}, are used.
 The effective thermal electron mass $m_{th}(300~\text{K}) = 0.278 \cdot m_e$, with the Boltzmann constant $k_B$ and the free electron mass $m_e$.

\begin{figure}[htpb]
\centering
\includegraphics[width=0.5\linewidth,angle=0]{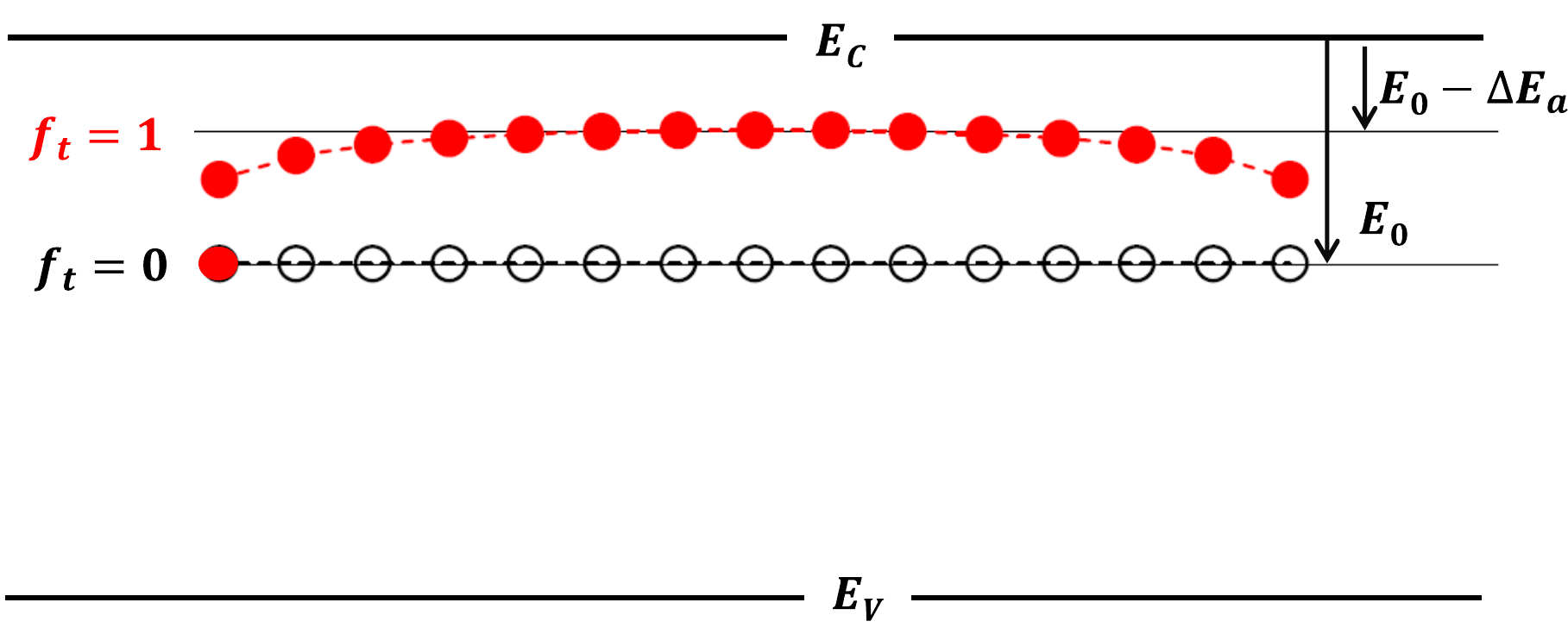}
 \caption{Schematic presentation of the potential energy of the electrons in a cluster consisting of 15 point defects equally spaced on a straight line.
 $E_V$ denotes the edge of the valence band and $E_C$ of the conduction band.
 The filled dots show the ionisation energies for the individual defects, when all traps are filled.
 The electron occupying the central defect has the lowest ionisation energy, $E_0 - \Delta E_a$.
 It will discharge first, when the sample is heated in the TSC~measurement.
 The open circles correspond to the ionisation energy when only one state is filled with an electron.
 Its value is $E_0$, the ionisation energy of the point defects, which build the cluster.}
\label{schema}
\end{figure}

  For point defects a constant value for $E_a$ is expected.
 As proposed in~\cite{scheinemann}, cluster defects can be characterized by an occupation-dependent value of $E_a (f_t)$.
  For the dependence $E_a (f_t)$, i.\,e.\ the change of the ionisation energy with the fraction $f_t$ of defects filled in a cluster, a linear dependence is assumed:
\begin{equation}\label{eaft}
  E_a(f_t) = E_0 - \Delta E_a \cdot f_t.
\end{equation}
 The schematic drawing of figure~\ref{schema} shows the potential energies of electrons in a cluster of 15 equally spaced electron traps on a straight line.
 The full dots correspond to $f_t = 1$, when all 15 traps are filled with electrons.
 The empty circles is the situation $f_t \rightarrow 0$, when only a single trap is filled and the energy is equal to the energy of the single isolated trap.
 The ionisation energy  $E_a = E_0 - \Delta E_a$ is minimal for $f_t = 1$, and reaches its maximum $E_a = E_0$, the value for point defects, for $f_t \rightarrow 0$.

\begin{figure}[htpb]
\centering
 \includegraphics[width=0.5\linewidth]{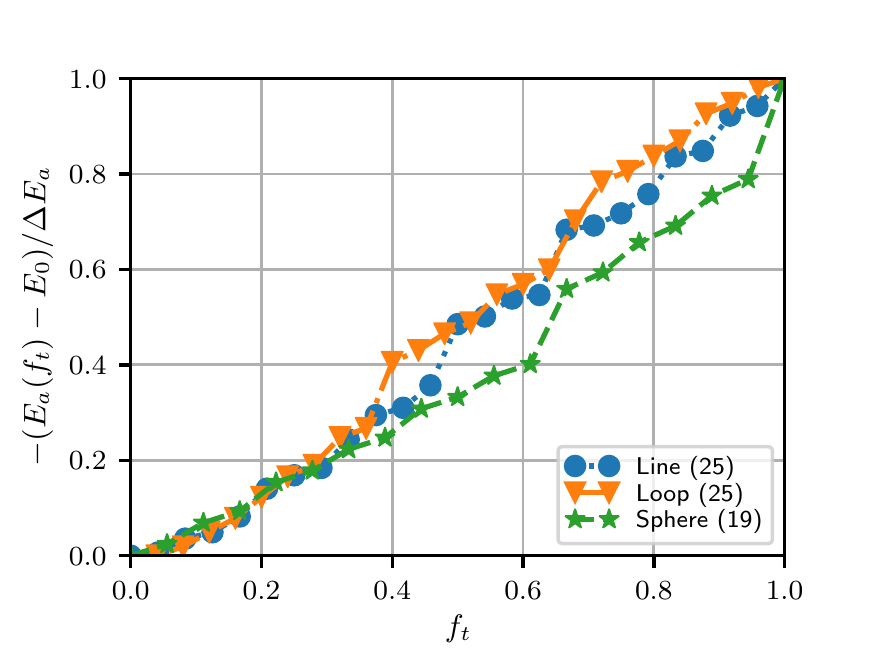}
\caption{Simulated dependence of $E_a(f_t)$ for different cluster topologies.
 Line~(25): 25 equally spaced point charges on a line,
 Loop~(25): 25 equally spaced point charges on a circle, and
 Sphere~(19): 19 point charges in a sphere on an equally spaced 3D-grid.}
\label{DeltaE}
\end{figure}

 Figure~\ref{DeltaE} shows the results of simple, electrostatic simulations of the dependence of $E_a(f_t)$.
 For \emph{Line}\,(25) the cluster is assumed to be a straight line of 25 equally spaced point defects, fixed in space.
 The energy difference between the conduction band and a single point defect is $E_0$.
 In the initial state, $f_t = 1$, each point defect is occupied by an electron and negatively charged.
 The point defect with the highest (negative) potential energy, $- \Delta E_a$, and thus the lowest $E_a$ value, $E_a = E_0 - \Delta E_a$, is discharged first, when the temperature is raised in the TSC measurement.
 Next, the point defect with the lowest $E_a$ value for the new charge configuration is found and discharged.
 This chain is continued, until all defects are discharged.
 The cluster \emph{Loop}\,(25) consists of 25 equally spaced point defects on a circle, and \emph{Sphere}\,(19) of 19 point defects in a sphere on a 3-D Cartesian grid.
 Figure~\ref{DeltaE} shows  that to a good approximation for all three topologies, $E_a(f_t)$ is linear in $f_t$.
 Loop clusters are discussed in \cite{scheinemann}, which actually triggered our studies.


 The data analysis consists in subtracting from the measured TSC current the dark current, obtained from the TSC measurement without trap filling.
 Then $\chi ^2$-fits of eq.~\ref{srheq} for different $T$ intervals are performed.
 For the errors 1\,\% times the measured currents have been assumed, which takes into account the uncertainty of the current measurement as well as the accuracy of the temperature regulation of the TSC setup.
 The free parameters of the model are $N_t$, $E_0$, $\Delta E_a$ and $\sigma$ for each defect, where $N_t$ is the density of point defects per volume, which are filled in the experiment at $T_0$, i.\,e. the product of cluster density times the number of filled point defects in a cluster.


 \section{Results}
 \label{results}
  \subsection{The VO$_i$ defect}
  \label{sect:VOi}

\begin{figure}[b!]
\centering
  \includegraphics[width=0.5\linewidth]{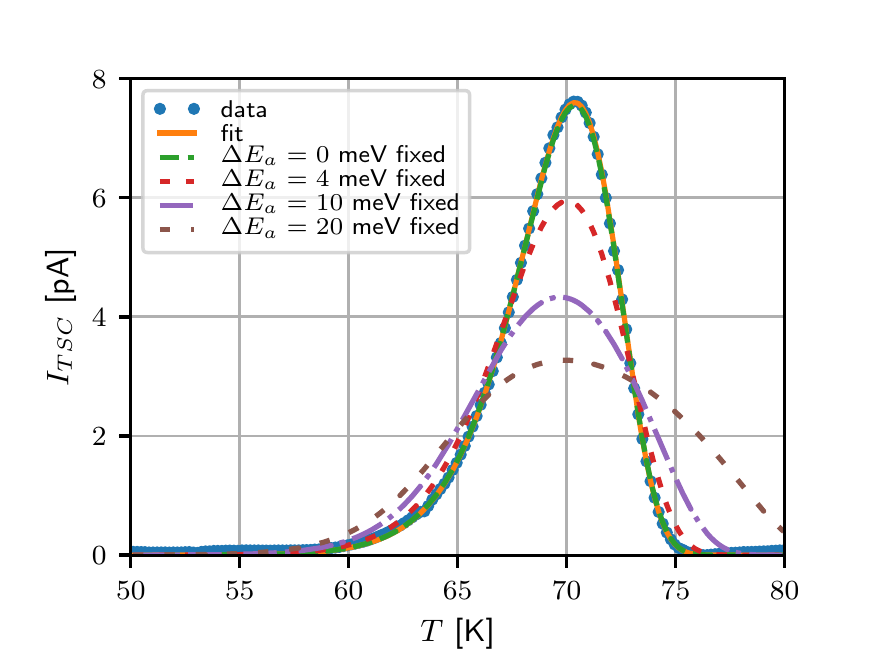}
  \caption{The measured TSC current after light injection at $T_0 = 10$~K (dots) for the point-like vacancy-oxygen defect (VO$_i$) after electron irradiation with $E_e = 27$~MeV to the fluence $\Phi=4.3\times10^{14}$~cm$^{-2}$.
  For the fit (solid line) in the range $T = 64.5$~K to 74.0~K, the activation energy $E_0$, $N_t$, $\sigma$, and $\Delta E_a$ are free parameters.
  The values $E_0 = 166.1 $\,meV and $\Delta E_a = -0.1 $~meV are found.
  For the other fits $E_0$ is set to 166.1\,meV, $\Delta E_a $ to the values given in the insert, and $N_t$ and $\sigma$ are free parameters.}
 \label{plot_voi}
\end{figure}

 As a first step, the vacancy-oxygen (VO$_i$) defect located at approximately $T = 70$~K is investigated.
 The VO$_i$ is known to be a point defect.
 According to reference~\cite{moll} it is an electron trap with an energy $E_0 = 176$\,meV from the conduction band and an electron capture cross-section $\sigma = 79 \times 10^{-16}$~cm$^{-2}$.
 The values in reference\,\cite{brotherton} are $E_0 = 169$\,meV and $\sigma =(100 \pm 10)\times 10^{-16}$\,cm$^{-2}$.
 In figure\,\ref{plot_voi} the TSC data after electron irradiation with $E_e= 27$~MeV to the fluence $\Phi = 4.3 \times 10^{14}$~cm$^{-2}$ and 30~min annealing at $80^{\,\circ }$C are shown as dots.
 The fit results are shown as a solid line, and the parameters are given in table~\ref{tab:VOi}.
 For the assumed 1\,\% uncertainty for the $I_{TSC}$ measurement, a $\chi ^2 =25$ for $NDF = 47$ degrees of freedom is found.
 We note that the differences \emph{fit -- measured} are dominated by the systematics of the temperature regulation and not by the statistical fluctuations of the current measurement.
 The values $E_0 = 166.1 \pm 0.7$\,meV and $\sigma = (20 \pm 2 ) \times 10^{-16}$\,cm$^{-2}$ (statistical errors only) differ significantly from the values given in \cite{moll}.
 If $E_0 = 176$\,meV is assumed, the $\chi ^2$ increases to 61 and $\sigma = 80 \times 10^{-16}$\,cm$^{2}$ is obtained. However, the value of $\Delta E_a = 0.88 \pm 0.03$\,meV is incompatible with zero, the value expected for a point defect.
 For the remaining curves of figure\,\ref{plot_voi} the value $E_0 = 166.1 $\,meV is assumed and $\Delta E_a $ is fixed to the values given in the insert.
 Values significantly larger than zero are incompatible with the data.
 We conclude that $\Delta E_a$ values, which differ significantly from zero are excluded.


\begin{table}[tb!]
 \centering
	\begin{tabular}{ccccc}
	   \toprule
		$E_0$ [meV] & $\Delta E_a$ [meV] & $N_t/10^{12}$ [cm$^{-3}$] &
		$\sigma /10^{-16}$ [cm$^2$] & $\chi ^2 / NDF$\\
	    \midrule
		$166.1 \pm 0.7$& $ -0.1 \pm 0.1$ & $3.44 \pm 0.01$  & $19.7 \pm 2.2$ & 25/47 \\
		 176           & $0.88 \pm 0.03$ & $3.43 \pm 0.01$  & $80.0 \pm 0.2$ & 61/48 \\
\bottomrule	
	\end{tabular}
   \caption{Results of the fits to $I_{TSC}$ in the temperature range 64.5 to 74.0~K for the diode irradiated to the fluence of $\Phi = 4.3 \times 10^{14}$~cm$^{-2}$ by electrons with $E_e = 27$~MeV.
   The first row gives the results for the fit with $E_0$ left free, and the second line the fit with $E_0$ constrained to the value from reference~\cite{moll}.}
 \label{tab:VOi}	
\end{table}

 The data after irradiation by electrons with energies $E_e$ between 3.5 and 15~MeV and after 30 min annealing at 80$^{\,\circ}$C, have also been analysed.
 In all cases, $\Delta E_a$ values close to zero and consistent values for $E_0$ and $\sigma$ are obtained.
 This confirms that VO$_i$ is a point defect, and validates the analysis method for an isolated point defect.

\subsection{Cluster defects}
 \label{sect:cluster}

\begin{figure}[!ht]
  \centering
   \begin{subfigure}[a]{0.5\linewidth}
    \includegraphics[width=1\linewidth]{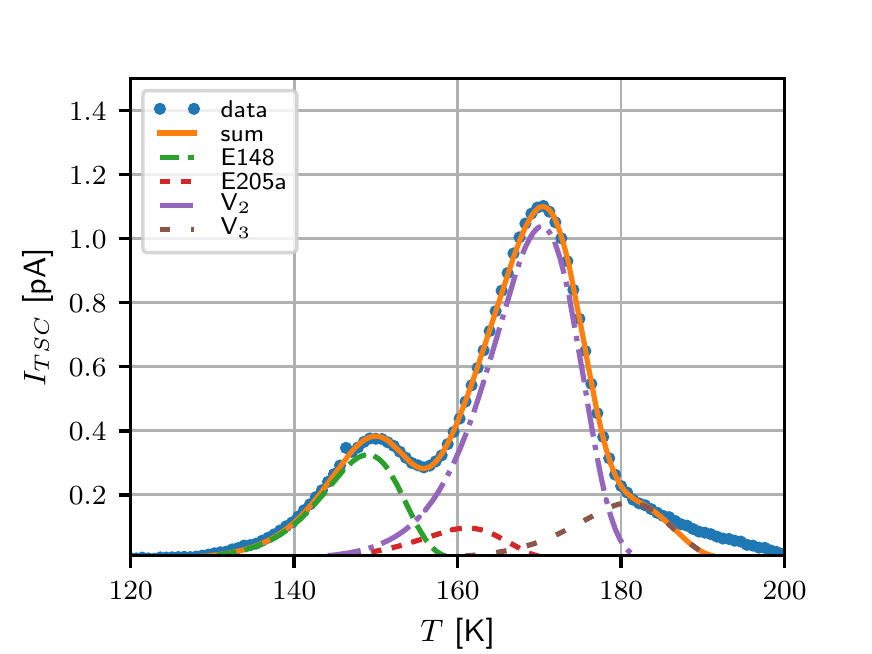}
    \caption{ }
     \label{fig:Ee3p5}
   \end{subfigure}%
    ~
   \begin{subfigure}[a]{0.5\linewidth}
    \includegraphics[width=1\linewidth]{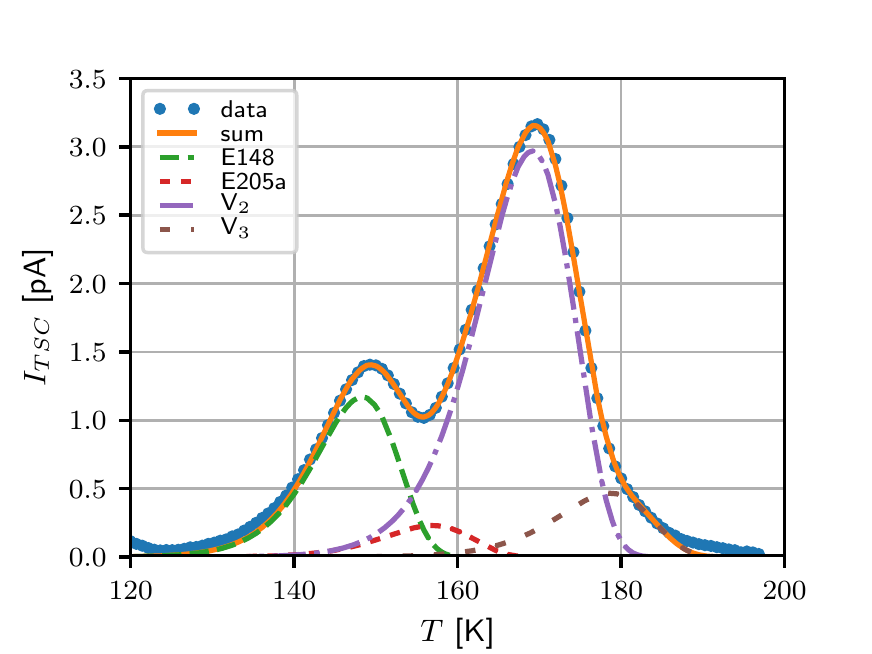}
    \caption{ }
     \label{fig:Ee6}
   \end{subfigure}
    ~
   \begin{subfigure}[a]{0.5\linewidth}
    \includegraphics[width=1\linewidth]{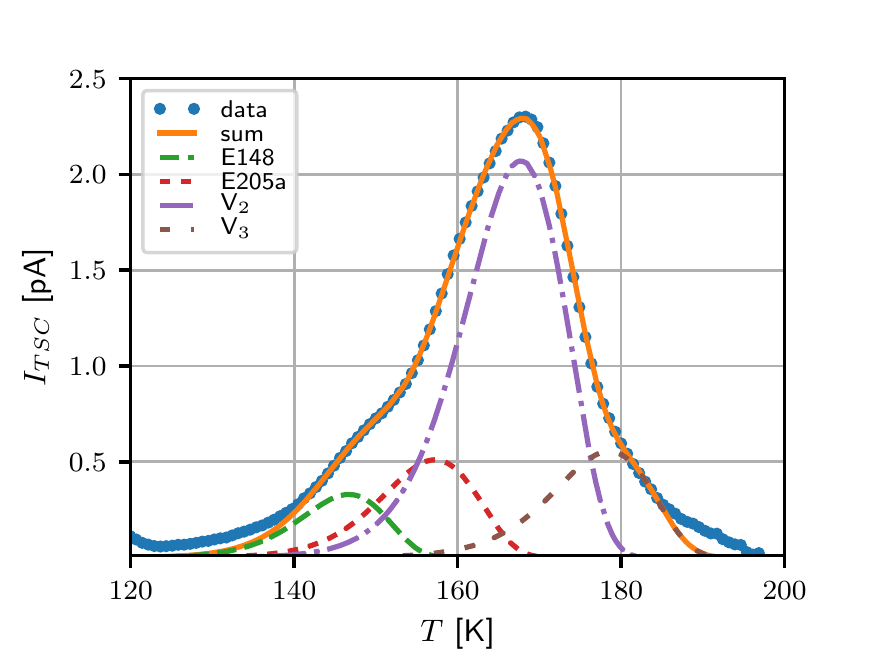}
    \caption{ }
     \label{fig:Ee15}
   \end{subfigure}%
    ~
   \begin{subfigure}[a]{0.5\linewidth}
    \includegraphics[width=1\linewidth]{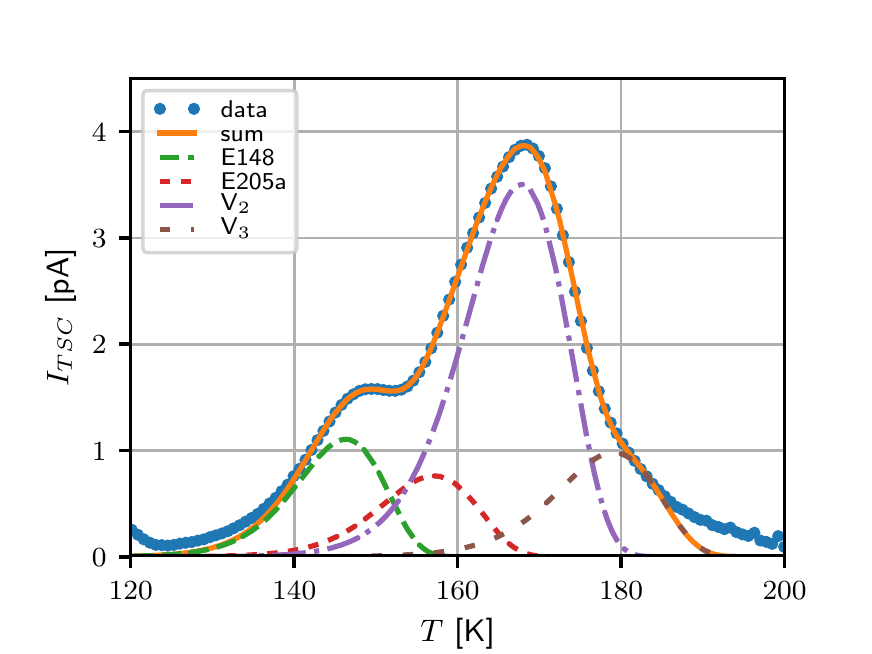}
    \caption{ }
     \label{fig:Ee27}
   \end{subfigure}%
   \caption{\,TSC spectra after 30~min annealing at $80^{\,\circ}$C with the fits described in the text for the temperature interval $T = 144.1 - 184.6$~K and irradiations with electrons of energies
    (a) 3.5~MeV,
    (b) 6~MeV,
    (c) 15~MeV, and
    (d) 27~MeV.}
  \label{fig:Ee}
 \end{figure}

 Next, the region $T = 120 - 200$~K is investigated, where overlapping defect clusters are located, which are responsible for part of the generation current, charge trapping and changes in the effective space-charge density\,\cite{radu2}.
 The TSC~spectra for the irradiations at the different $E_e$~values after 30~min annealing at 80$^{\,\circ}$C are shown in fig.~\ref{fig:Ee}.


 To describe the data, four defects have been assumed: the di-vacancy V$_2$, the tri-vacancy V$_3$ and two additional defects, E148 and E205a.
 The values of $E_0$ from the literature are given in table \ref{tab:Cluster}.
 Using eq.~\ref{srheq} the sums of the contributions of all four defects have been fitted to the $I_{TSC}$ measurements.
 In the fit, the literature values for $E_0$, a $\Delta E_a$\,value common for the two defects E148 and V$_3$, and another $\Delta E_a$\,value common for E205a and V$_2$, have been assumed.
 Other assumptions, e.\,g. a common $\Delta E_a$\,value for all 4 states, gave qualitatively similar results.
 Without such constraints, which are only poorly motivated, the fits do not converge to stable results.
 Figure~\ref{fig:Ee} compares the fit results to the data, and table~\ref{tab:Cluster} shows as an example the parameters derived for the $E_e = 15$~MeV data.
 The TSC\,spectra are well described by the fits.
 It has been verified that they  cannot be described with acceptable trap parameters if $\Delta E_a = 0$ is assumed,
 and conclude that the  model with an occupation-dependent ionisation energy provides a good description of the TSC~spectra of the cluster defects.

 \begin{table}[bt!]
 \centering
	\begin{tabular}{cccccc}
	   \toprule
		Defect & $E_0$ [meV]& Ref. & $\Delta E_a$ [meV] & $N_t$/$10^{12}$ [cm$^{-3}]$ & $\sigma $/$10^{-16}$ [cm$^2$] \vspace{0.5mm} \\
	    \midrule
		E148         & 359 & \cite{junkes, pinti2001}& 4.3  & 0.33  & 4.6 \\
		E205a        & 393 & \cite{junkes, moll} & 7.6  & 0.56  & 6.1 \\
		V$_2$($-/0$) & 424 & \cite{moll}         & 7.6  & 2.44  & 7.0 \\
		V$_3$($-/0$) & 456 & \cite{radu2, moll}  & 4.3  & 0.65  & 9.7 \\
	  \bottomrule	
	\end{tabular}
	\caption{Parameters for the four electron traps in the $T = 120 - 200$~K region:
   $E_0$ is the energy distance of the isolated point defects from the conduction band taken from the reference given in the third column.
   $\Delta E_a$, $N_t$ and $\sigma $ are the cluster-related energy shifts, the density of traps filled at $T_0$ and the electron cross-sections obtained from the fit for $T = 144.1 - 184.6$~K of eq.~\ref{srheq} to the data for $E_e = 15$~MeV after 30~min annealing  at $80^{\,\circ} $C.
   The statistical errors are about 0.1\,meV for $\Delta E_a $ and at the few\,\% level for $N_t$ and $\sigma $.
   The systematic uncertainties, e.\,g. from the choice of the fit interval, are significantly larger.}
 \label{tab:Cluster}	
\end{table}

\begin{figure}[!ht]
  \centering
   \begin{subfigure}[a]{0.5\linewidth}
    \includegraphics[width=1\linewidth]{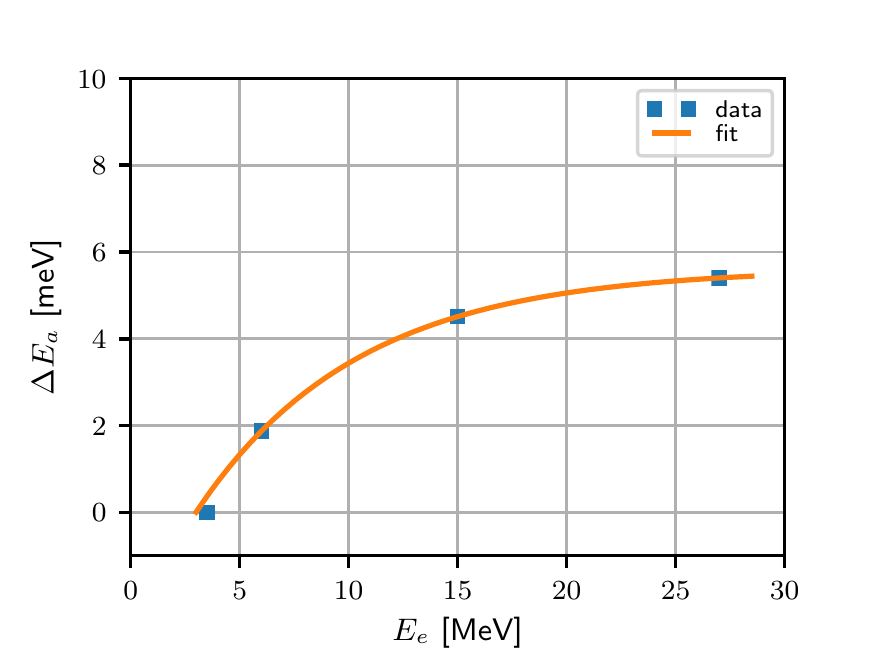}
    \caption{ }
     \label{fig:DE148Ee}
   \end{subfigure}%
    ~
   \begin{subfigure}[a]{0.5\linewidth}
    \includegraphics[width=1\linewidth]{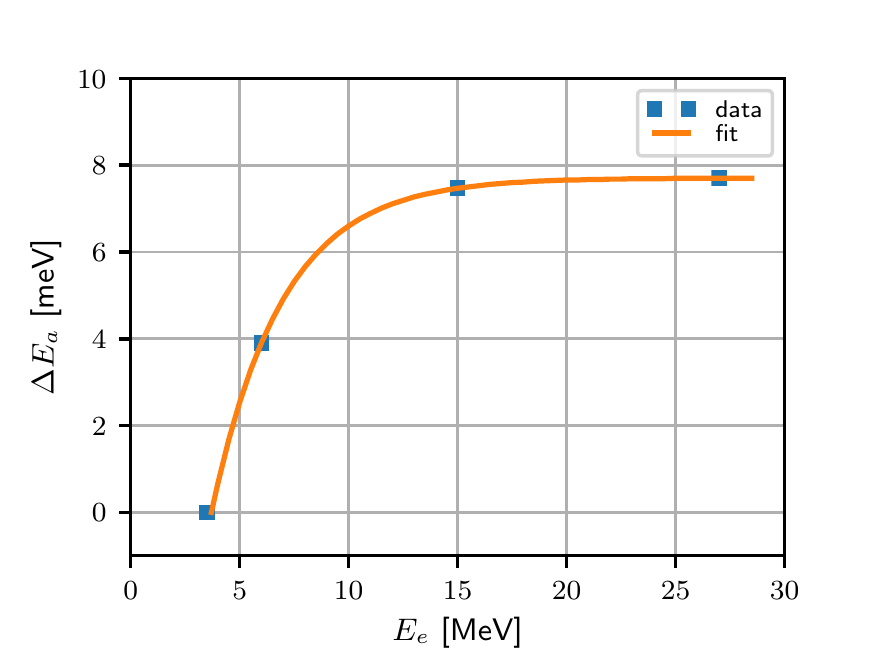}
    \caption{ }
     \label{fig:DEV2}
   \end{subfigure}
   \caption{\,$\Delta E_a$ after 30~min annealing at $80^{\,\circ} $C as a function of electron energy, $E_e$, for
    (a) the E148/V$_3$-, and
    (b) the E205a/V$_2$-defect.}
  \label{fig:DEaEe}
 \end{figure}

 In order to investigate the dependence of the cluster formation on the electron energy, $E_e$, figure\,\ref{fig:DEaEe} shows $\Delta E_a (E_e)$ for the V$_3$/E148 and the V$_2$/E205a clusters obtained from the fits described above.
 The $\Delta E_a$~values for $E_e = 3.5$~MeV are compatible with zero.
 For higher electron energies the $\Delta E_a$ values differ  significantly from zero and the energy dependence is as expected for a threshold of cluster formation in the region $E_e = 3.5 - 6 $\,MeV.

 The phenomenological function with the free parameters $B$, $\gamma _e$ and $E_{th}$,
 \begin{equation}\label{eqea}
  \Delta E_a (E_e) = B \cdot \left[ 1 - \mathrm{exp} \left( \frac{E_e - E_{th}}{\gamma_e} \right) \right],
 \end{equation}
 is fit to the $\Delta E_a(E_e)$\,data.
 For the threshold values
 $E_{th}^{\mathrm{V}_3/\mathrm{E148}} = 3.0\pm 0.5$~MeV and
 $E_{th}^{\mathrm{V}_2/\mathrm{E205a}} = 3.7\pm 0.5$~MeV are obtained.

 We note that for the models presented in section \ref{sect:method}, a value of $\Delta E_a = 7.5 $~meV corresponds to a spacing of about 100~nm of the point defects, which are filled at $T_0 = 10$\,K in the clusters, to be compared to the silicon-lattice constant of 0.543\,nm at 300\,K.
 The values are compatible with simulations of electron-induced damage in silicon, which take into account the Mott cross-section, the vacancy and interstitial generation, as well as their diffusion and recombination~\cite{lindstroem}.

\subsection{Cluster annealing}
 \label{sect:anneal}

\begin{figure}[!ht]
  \centering
   \begin{subfigure}[a]{0.5\linewidth}
    \includegraphics[width=1\linewidth]{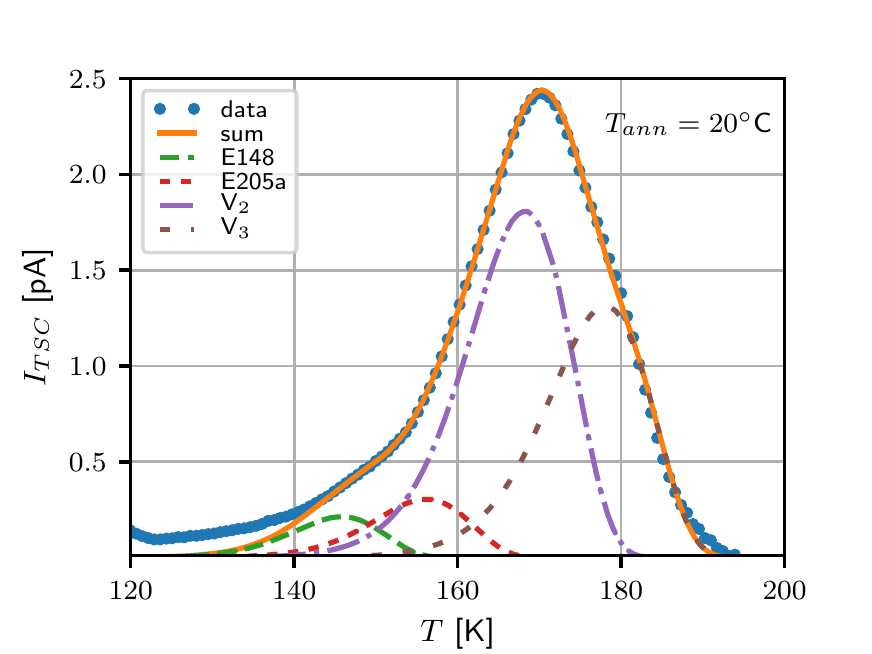}
    \caption{ }
     \label{fig:Tann20}
   \end{subfigure}%
    ~
   \begin{subfigure}[a]{0.5\linewidth}
    \includegraphics[width=1\linewidth]{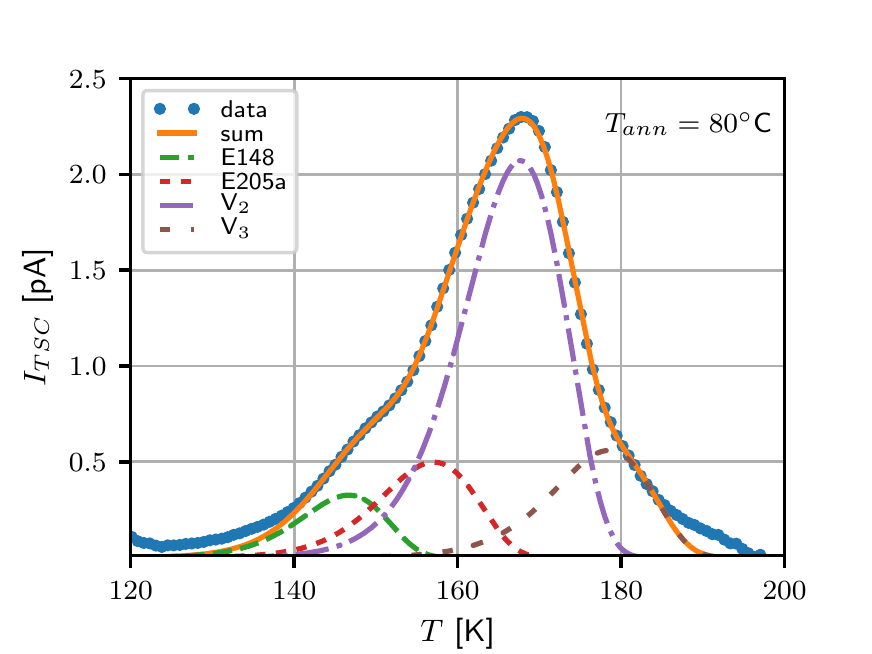}
    \caption{ }
     \label{fig:Tann80}
   \end{subfigure}
    ~
   \begin{subfigure}[a]{0.5\linewidth}
    \includegraphics[width=1\linewidth]{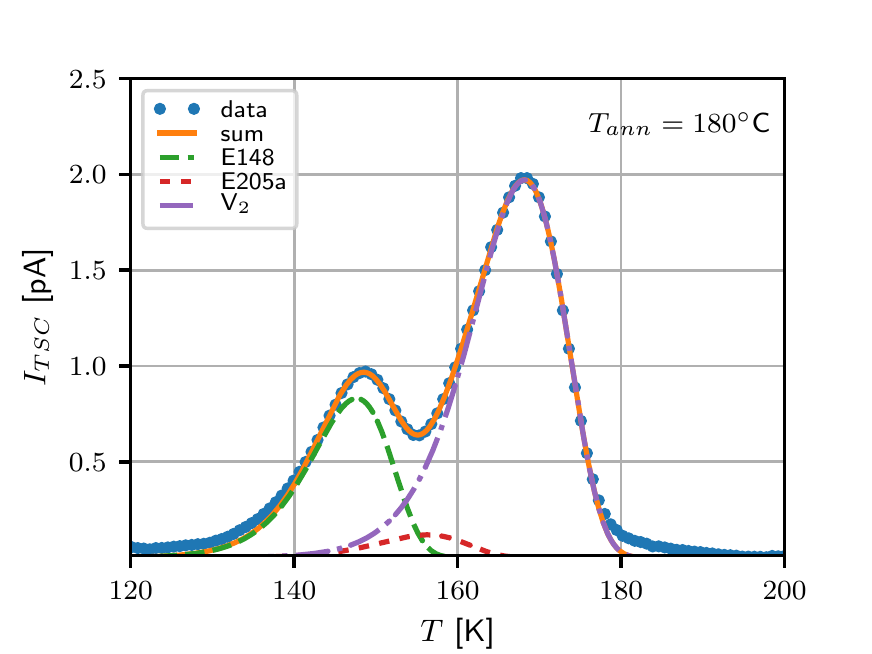}
    \caption{ }
     \label{fig:Tann180}
   \end{subfigure}%
    ~
   \begin{subfigure}[a]{0.5\linewidth}
    \includegraphics[width=1\linewidth]{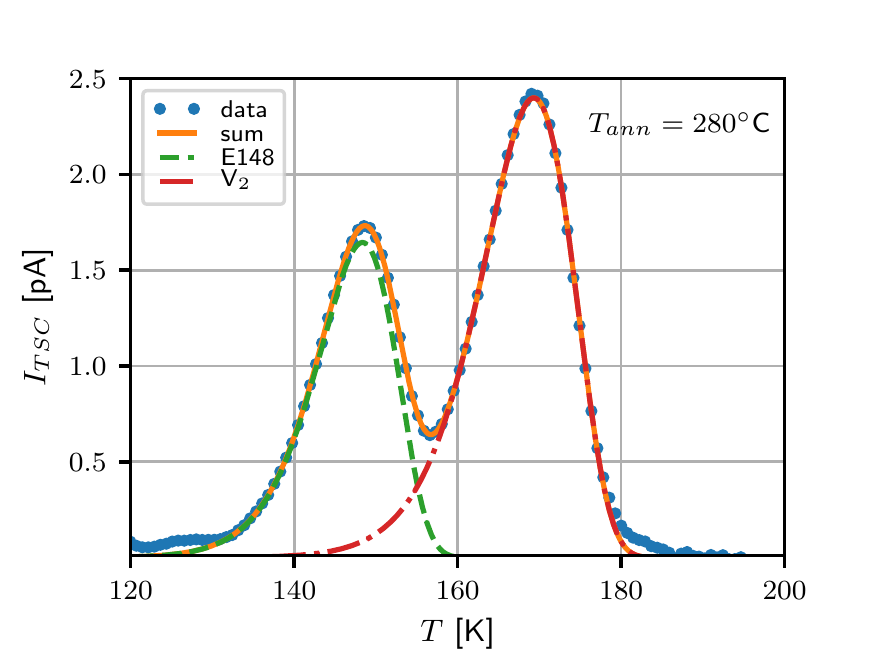}
    \caption{ }
     \label{fig:Tann280}
   \end{subfigure}%
   \caption{Measured TSC spectra, $I_{TSC}(T)$, and results of the fits assuming four defect clusters in the $120 - 200$~K region for the diode irradiated by 15~MeV electrons to a fluence of 2.6$\cdot$10$^{14}$~cm$^{-2}$ for
    (a) \emph{as irradiated}, and after 30~min annealing at
    (b) 80$^{\,\circ}$C,
    (c) 180$^{\,\circ}$C, and
    (d) 280$^{\,\circ}$C.}
  \label{fig:Tann}
 \end{figure}

 In the following, the analysis of the data for the irradiation with $E_e = 15$\,MeV for the conditions \emph{as irradiated} and after isochronal annealing for $T_{ann} = 80 - 280^{\,\circ}$C in 20$^{\,\circ}$C steps and 30~min annealing time, is presented.
 The temperature history for \emph{as irradiated} is only approximately known:
 After the 15~MeV irradiation at Dresden-Rossendorf the diode had been at room temperature for about eight hours during the transport to Hamburg, where it has been stored in the freezer at $-25^{\,\circ}$C.
 In the following this condition is labeled $T_{ann} = 20^{\,\circ}$C.
 In figure\,\ref{fig:Tann} we show as examples the comparison of the fit results with the measurements for \emph{as irradiated} and after the isochronal annealing steps of 80$^{\,\circ}$C, 180$^{\,\circ}$C and 280$^{\,\circ}$C.
 For the fits the assumptions discussed in section~\ref{sect:cluster} are made.
 For all $T_{ann}$\,values the data are well described by the model.

\begin{figure}[!ht]
  \centering
   \begin{subfigure}[a]{0.5\linewidth}
    \includegraphics[width=1\linewidth]{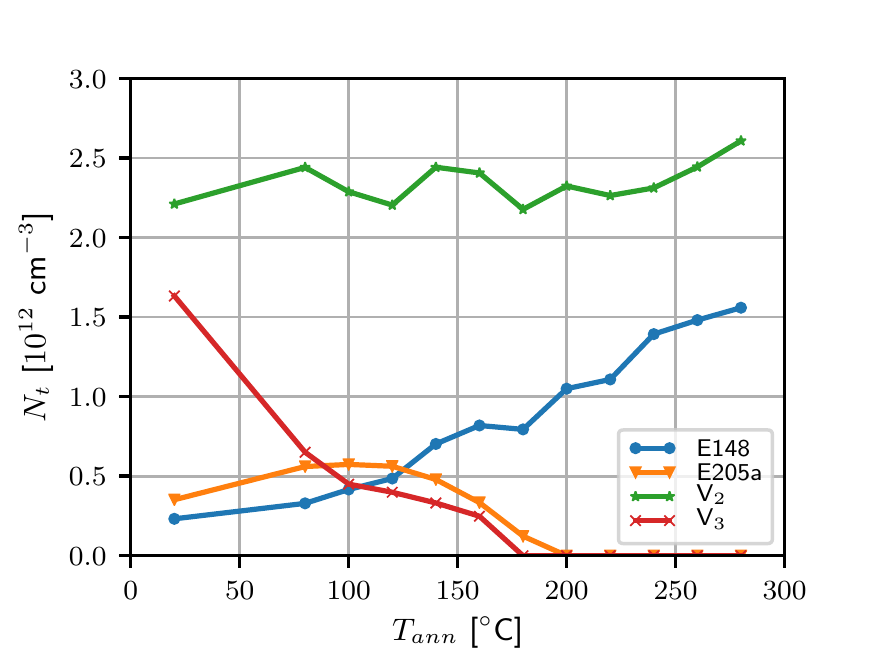}
    \caption{ }
     \label{fig:NtTann}
   \end{subfigure}%
    ~
   \begin{subfigure}[a]{0.5\linewidth}
    \includegraphics[width=1\linewidth]{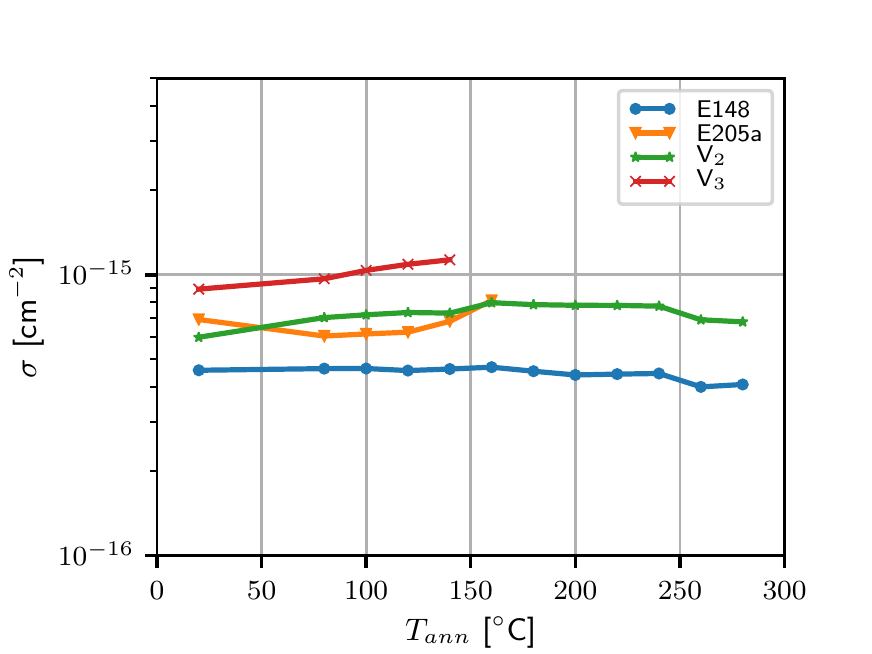}
    \caption{ }
     \label{fig:SigmaTann}
   \end{subfigure}
   \caption{ (a) Densities of traps filled at $T_0$, $N_t$, and (b) electron cross-sections, $\sigma $, for the diodes irradiated by 15~MeV electrons to a fluence of $2.6 \times 10^{14}$~cm$^{-2}$ as a function of the annealing temperature, $T_{ann}$.}
  \label{fig:ResultsTann}
 \end{figure}

 The results for $N_t$ of the four defects, and the cross-sections for electrons, $\sigma $, as a function of $T_{ann}$ are shown in figure\,\ref{fig:ResultsTann}.
  As discussed in section\,\ref{sect:method}, $N_t$ is not the density of filled clusters, but the product of the cluster density times the number of filled traps in the cluster.
 We observe that with increasing annealing temperature $N_t$ remains constant for V$_2$, increases for E148, and vanishes for V$_3$ and for E205a. More details on the annealing behavior of $V_3$ and E205a can be found in~\cite{junkes}.
 As expected the electron cross-sections do not depend on $T_{ann}$.

\begin{figure}[!bt]
  \centering
   \begin{subfigure}[a]{0.5\linewidth}
    \includegraphics[width=1\linewidth]{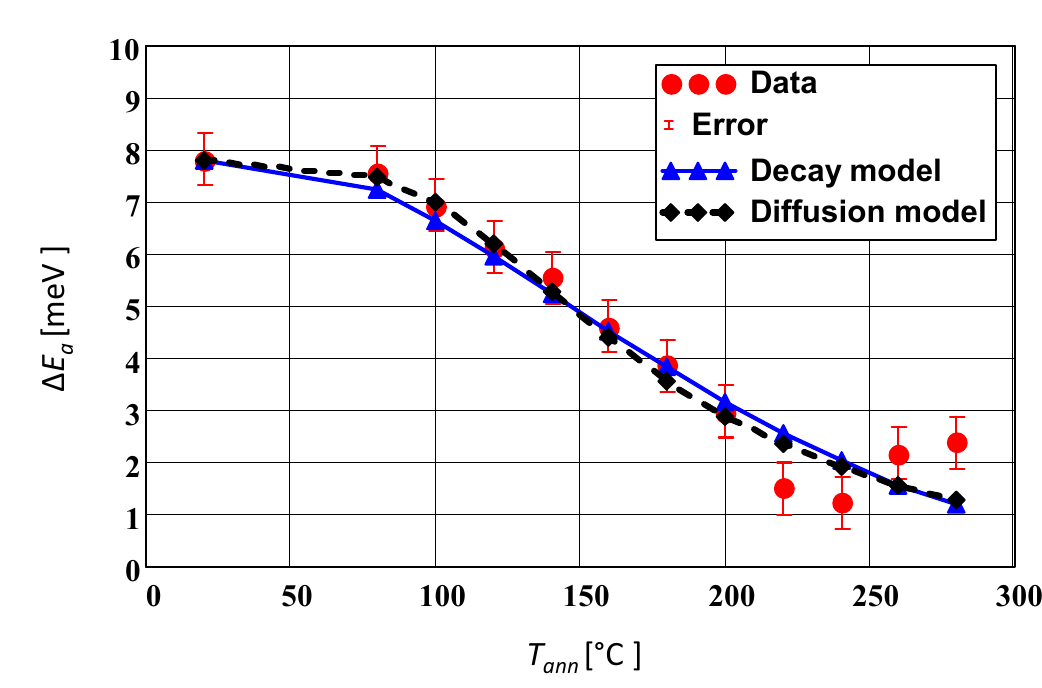}
    \caption{ }
     \label{fig:DEaV2Tann}
   \end{subfigure}%
    ~
   \begin{subfigure}[a]{0.5\linewidth}
    \includegraphics[width=1\linewidth]{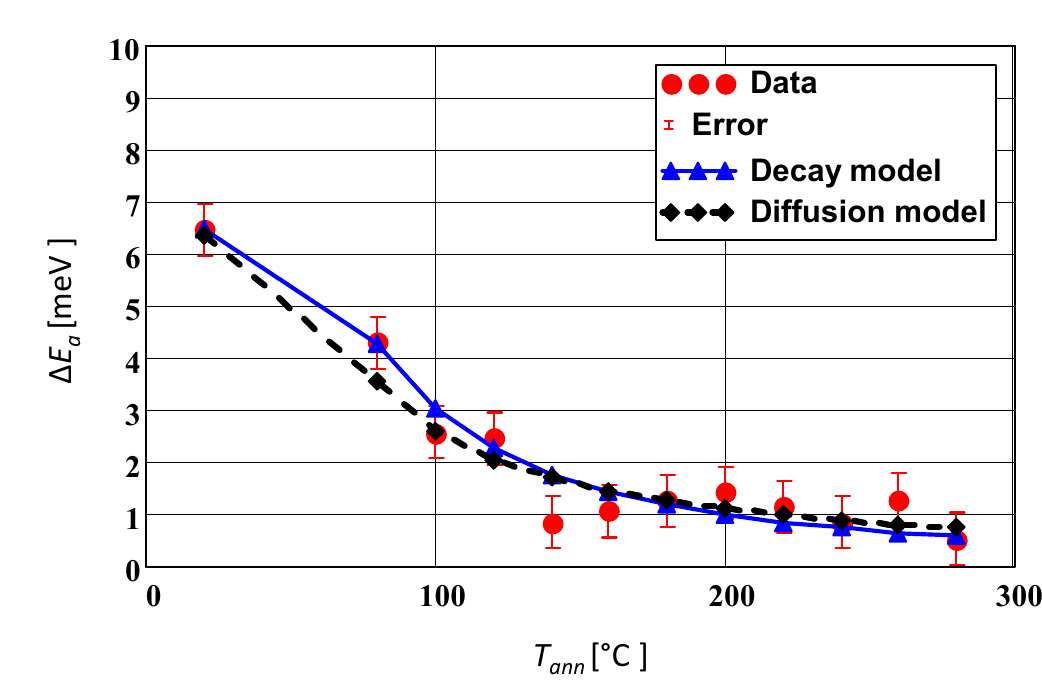}
    \caption{ }
     \label{fig:DEaV3Tann}
   \end{subfigure}
   \caption{ Dependence of $\Delta E_a (T_{ann})$ for the diode irradiated by 15~MeV electrons to a fluence of $2.6 \times 10^{14}$~cm$^{-2}$ for \emph{as irradiated}, plotted at $T_{ann} = 20^{\,\circ}$C, and after isothermal annealing at the temperature $T_{ann}$ for 30~min.
   The points with error bars are the data, the solid lines the fits by the \emph{first order decay model}, and the dashed lines the fits by the \emph{diffusion model} for
   (a) V$_2$/E205a and
   (b) V$_3$/E148.}
  \label{fig:DEaTann}
 \end{figure}

 The results for $\Delta E_a$ as a function of $T_{ann}$ for the traps V$_2$/E205a and V$_3$/E148 are shown in figure\,\ref{fig:DEaTann}.
 For V$_2$/E205a the value of $\Delta E_a$ remains nearly constant ($\approx 7.5$~meV) for the annealing at $80^{\,\circ}$C, then decreases to $\approx 2$~meV above $250^{\,\circ}$C.

 For V$_3$/E148 the decrease of $\Delta E_a$ with $T_{ann}$ is significantly faster, and a constant value of $\approx 1$~meV is reached  at $T_{ann} = 140^{\,\circ}$C.
 For the description of $\Delta E_a (T_{ann} ) $ two models are introduced:
 a \emph{first order decay model}, and a \emph{ diffusion model}.

 The \emph{first order decay model} uses the following formula for $\Delta E_a$ after the $i$-th annealing step
 \begin{equation}\label{DEamodel1}
   \Delta E_a(T_{ann}^{(i)})=\Delta E_a(T_{ann}^{(i-1)}) \cdot e^{-k(T_{ann}^{(i)})\cdot t_{ann}}
   \hspace{5mm} \mathrm{with} \hspace{5mm} k(T_{ann})=k_0 \cdot e^{-E_A/(k_B T_{ann})},
 \end{equation}
 with the annealing time $t_{ann} = 30$\,min.
 The first order process is described by the frequency factor $k_0$ and the activation energy $E_A$.
 For $\Delta E_a(T_{ann}^{(0)})$ the \emph{as irradiated} data assigned to $T_{ann} = 20^{\,\circ}$C are used.
 The results of the fits to the data are shown in figure\,\ref{fig:DEaTann} and in table~\ref{tab:EFfit}.
 The model provides an acceptable description of the data.

 The \emph{diffusion model} follows a similar approach as the model discussed in section~\ref{sect:method}, which was used to derive the linear dependence of the ionisation energy on the fraction of occupied defects.
 Each cluster is assumed to consist of $n_{clu}$ point defects, which diffuse during the annealing steps.
 For the temperature dependence of the diffusion constant the standard parametrisation
 \begin{equation}\label{Diff}
   D(T) = D_0 \cdot e^{-E_D/(k_B T)}
 \end{equation}
 is used.
 The activation energy for the diffusion of the point defects in the cluster is $E_D$, and $D_0$ a constant.
 For the fit $n$ clusters with $n_{clu}$ point defects each, are generated.
 The position of the $j$-th point defect in the $k$-th cluster after the $(i-1)$-st annealing step is denoted by $\vec{x}_{j,k}^{~(i-1)}$.
 The position after the $i$-th annealing step is $\vec{x}_{j,k}^{~(i)} = \vec{x}_{j,k}^{~(i-1)} + \vec{\delta x}$, where $\vec{\delta x}$ is a vector of three Gauss-distributed random numbers with width $\sigma _D (T_{ann}^{(i)}) = \sqrt{2 ~ D(T_{ann}^{(i)}) \cdot t_{ann}}$.
 Assuming that all point defects are occupied by electrons, the individual potential energies $V_{j,k}^{~(i)} $ are calculated.
 The biggest $|V_{j,k}^{~(i)}| $ is assumed to be the $\Delta E_a$\,value of the $k$-th cluster, and the average $\Delta E_a$ of the $n$ clusters is fitted to the data.
 The initial positions of the point charges in the $\vec{x}_{j,k}^{~(0)}$ are generated as $n_{clu}$ points spaced by $d_0$ on a straight line each smeared with a vector of three Gauss-distributed random numbers with width $\sigma _0$.
 For a given number of clusters, $n$, and cluster size, $n_{clu}$, the free parameters of the fit are $\sigma _0$, $d_0$, $D_0$ and $E_D$.
 For the fits to the data shown in figure\,\ref{fig:DEaTann}, $n = 1000$ and $n_{clu} = 15 $ are used.
 Fits have also been performed for other $n_{clu}$\,values.
 The $\chi ^2$~values obtained depend only weakly on $n_{clu}$ and there is a shallow minimum around  $n_{clu} = 15 $.
 The parameters obtained from the fit are given in table~\ref{tab:RKfit}, where the diffusion constants  at 423.15\,K, $D_{423.15 \mathrm{K}} = D_0 \cdot e^{-E_D/(k_B \cdot 423.15\,\mathrm{K})}$, instead of $D_0$ are shown.
 Using the value of $D$ at the average annealing temperature 423.15\,K has the advantage that $D_{423.15 \mathrm{K}}$ and $E_D$ are essentially uncorrelated.
 In addition, it is more easily interpreted than $D_0$, which corresponds to the value of $D$ for $T_{ann} \rightarrow \infty$.

\begin{figure}[!bt]
  \centering
   \begin{subfigure}[a]{0.5\linewidth}
    \includegraphics[width=1\linewidth]{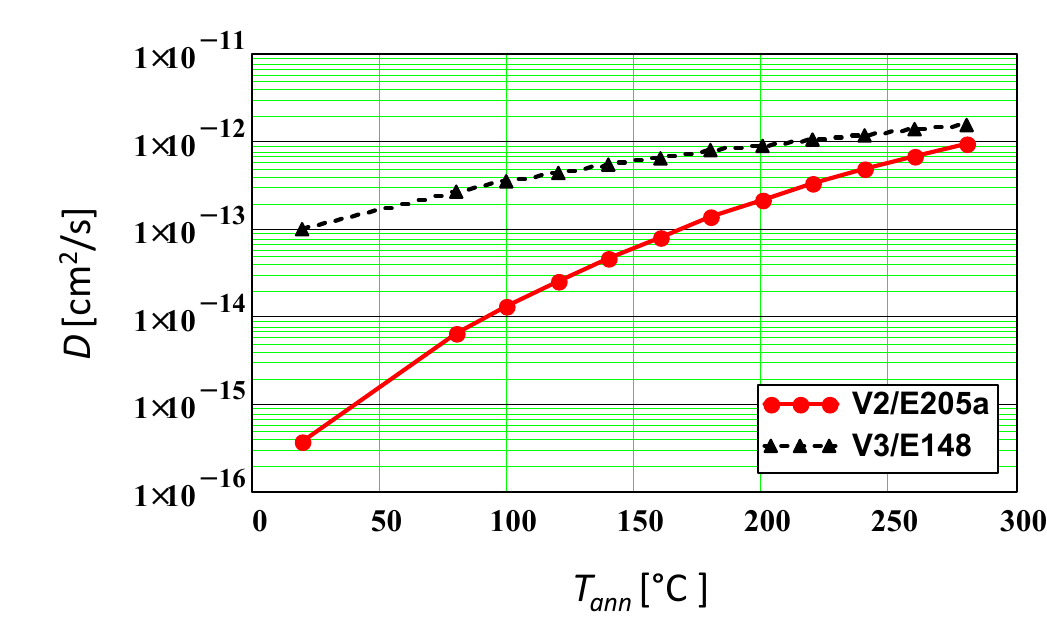}
    \caption{ }
     \label{fig:DTann}
   \end{subfigure}%
    ~
   \begin{subfigure}[a]{0.5\linewidth}
    \includegraphics[width=0.97\linewidth]{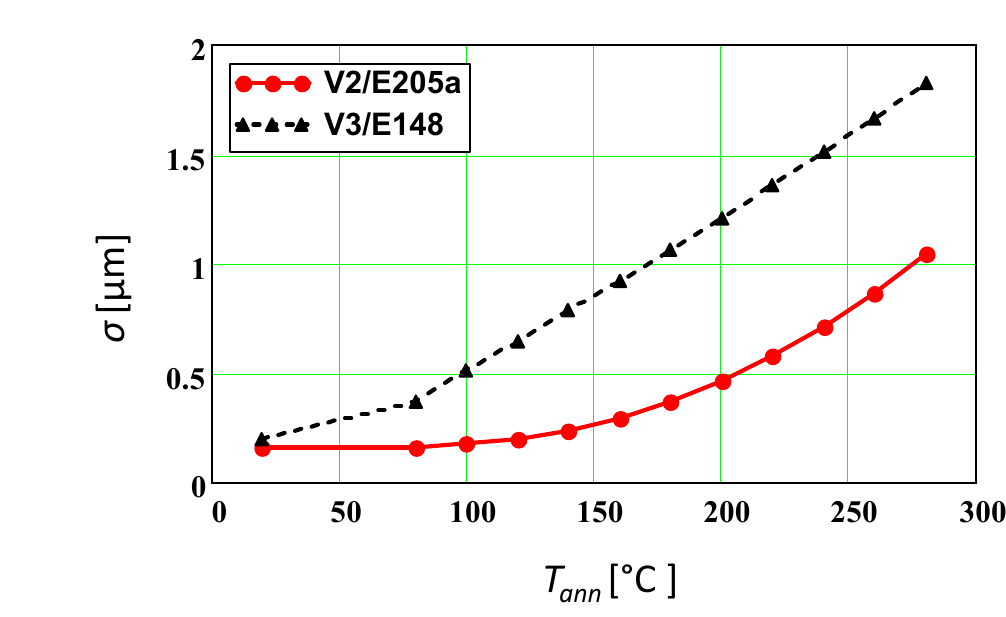}
    \caption{ }
     \label{fig:SigTann}
   \end{subfigure}
   \caption{ Results from the fits of the \emph{diffusion model} for clusters containing $n_{clu} = 15 $ point defects as a function of the temperature $T_{ann}$ of 30\,min isochronal annealing for the diode irradiated to a fluence of $2.6 \times 10^{14} $\,cm$^{-2}$.
   (a) Diffusion constant, and
   (b) $\sigma $~parameter of the Gaussian describing the diffusion of the point defects in the clusters.}
  \label{fig:DiffTann}
 \end{figure}

 We note that, whereas the values of the parameters given in table\,\ref{tab:RKfit} change significantly if the fit region is changed, the dependencies of $D(T_{ann})$ and $\sigma (T_{ann})$, shown in figure\,\ref{fig:DiffTann}, are much less affected.
 The values of the diffusion activation energy $E_D \approx 0.42 $\,eV for V$_2$/E205a is similar to the value given in\,\cite{myers} for the vacancies V$^0$ and V$^{-1}$, and $E_D \approx 0.15 $\,eV for V$_3$/E148 similar to the interstitial Si$_i^0$.


  \begin{table}[bt!]
 \centering
	\begin{tabular}{ccccc}
	   \toprule
		Defect & $k_0$ [1/s]& $ E_A$ [eV] & $\chi ^2/NDF$ & Fit range [$^\circ $C] \vspace{0.5mm} \\
	    \midrule
		V$_2$/E205a       & $1.6 \times 10^{-3}$ & 0.113  & 15.6/10 & $20 - 280$ \\
		V$_2$/E205a       & $42 \times 10^{-3}$  & 0.225  &  1.2/8  & $20 - 240$ \\
		V$_3$/E148        & $7.2 \times 10^{-6}$ & $-0.105$ &  7.9/10 & $20 - 280$ \\
		V$_3$/E148        & $3.6 \times 10^{-4}$ & 0.014  &  2.3/3  & $20 - 140$ \\
	  \bottomrule	
	\end{tabular}
	\caption{ Parameters obtained for the fit of the \emph{first order decay model}, equation\,\ref{DEamodel1}, to the $\Delta E_a (T_{ann})$ results shown in figure\,\ref{fig:DEaTann}. The last column gives the $T_{ann}$\,range of the fit}
 \label{tab:EFfit}	
\end{table}

  \begin{table}[bt!]
 \centering
	\begin{tabular}{ccccccc}
	   \toprule
		Defect & $D_{423.15 \mathrm{K}}$ [nm$^2$/s]& $ E_D$ [eV] & $d_0$ [nm] & $\sigma_0$ [nm] & $\chi ^2/NDF$ & $T_{ann} $ [$^\circ $C] \vspace{0.5mm} \\
	    \midrule
		V$_2$/E205a       &  6.3 & 0.423 & 27 & 159 & 6.4/8 & $20 - 280$ \\
		V$_3$/E148        &  60  & 0.146 & 27 & 200 & 5.7/8 & $20 - 280$ \\
	  \bottomrule	
	\end{tabular}
	\caption{ Parameters obtained for the \emph{diffusion model} described in the text, to the $\Delta E_a (T_{ann})$ results shown in figure\,\ref{fig:DEaTann}.}
 \label{tab:RKfit}	
\end{table}

\begin{figure}[!ht]
  \centering
   \begin{subfigure}[a]{0.5\linewidth}
    \includegraphics[width=0.8\linewidth]{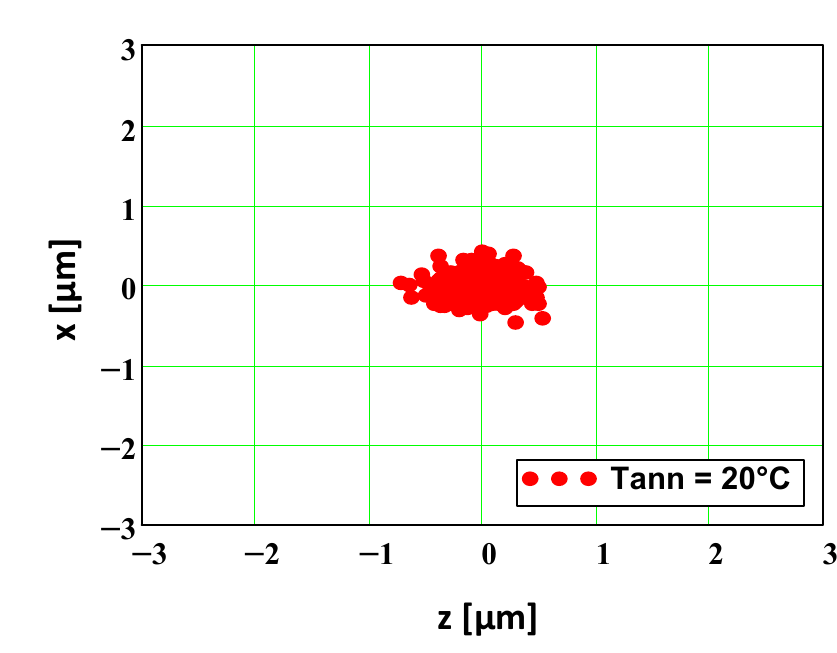}
    \caption{ }
     \label{fig:XZ20}
   \end{subfigure}%
    ~
   \begin{subfigure}[a]{0.5\linewidth}
    \includegraphics[width=0.8\linewidth]{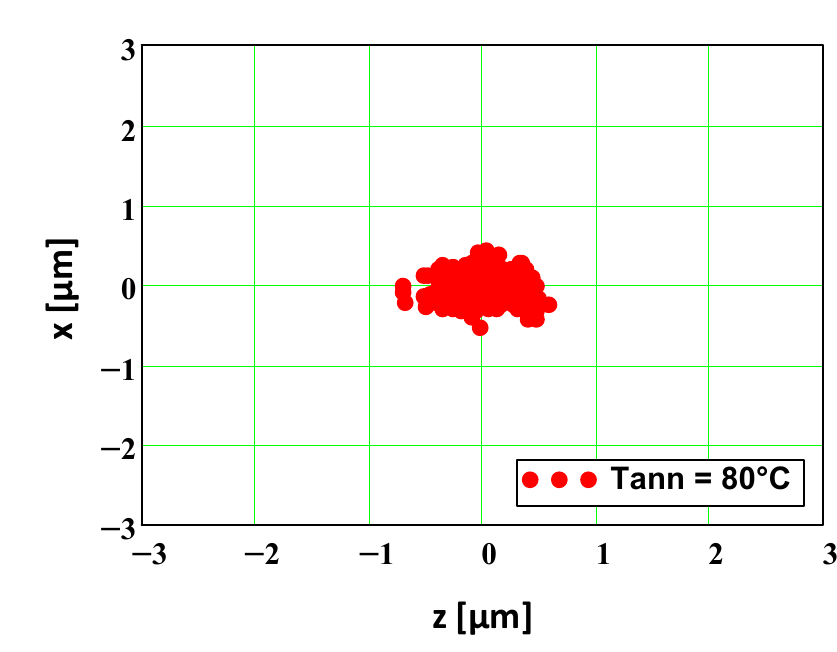}
    \caption{ }
     \label{fig:XZ80}
   \end{subfigure}
    ~
   \begin{subfigure}[a]{0.5\linewidth}
    \includegraphics[width=0.8\linewidth]{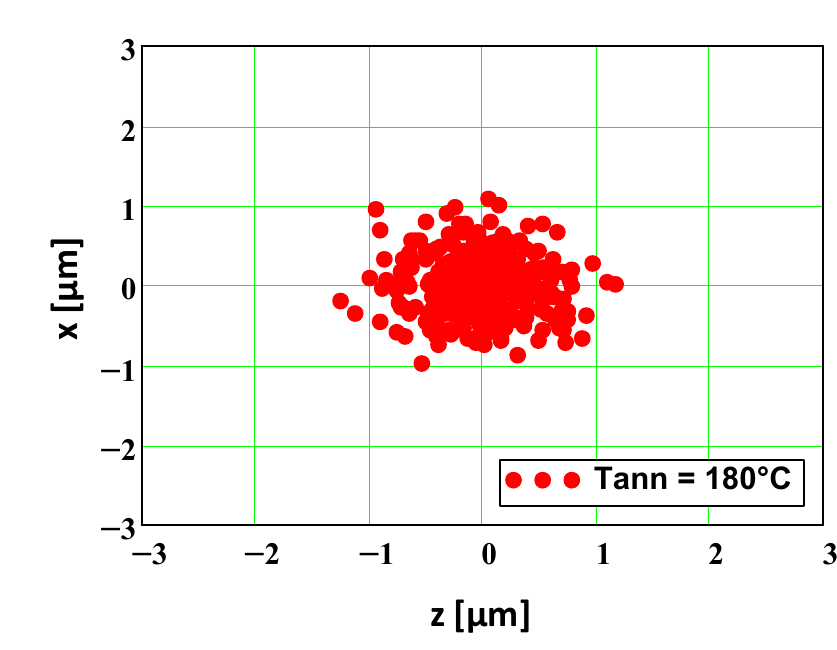}
    \caption{ }
     \label{fig:XZ180180}
   \end{subfigure}%
    ~
   \begin{subfigure}[a]{0.5\linewidth}
    \includegraphics[width=0.8\linewidth]{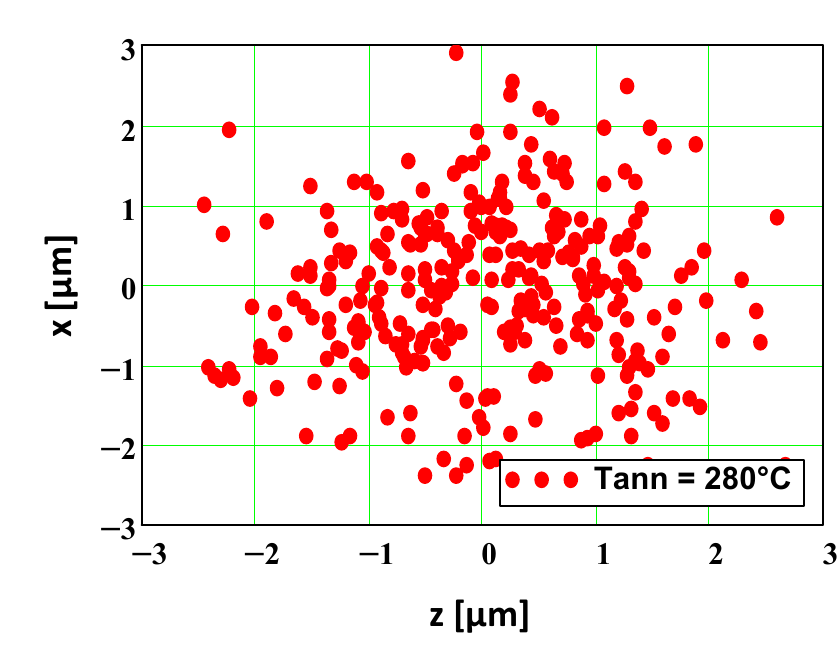}
    \caption{ }
     \label{fig:XZ280}
   \end{subfigure}%
   \caption{Diffusion of the point charges in the V$_2$/E205a cluster simulated using the parameters of table\,\ref{tab:RKfit} for
    (a) \emph{as irradiated}, and after 30~min annealing at
    (b) 80$^{\,\circ}$C,
    (c) 180$^{\,\circ}$C, and
    (d) 280$^{\,\circ}$C.
    To illustrate the diffusion, for the presentation 20 clusters of 15 point charges each are superposed.}
  \label{fig:XZTann}
 \end{figure}

 As a pictorial presentation of the diffusion of the point defects in the clusters, figure\,\ref{fig:XZTann} shows for the V$_2$/E205a cluster the $x - z$\,distribution of the superposition of 20 cluster with 15 point defects each, using the parameters of table\,\ref{tab:RKfit} for the conditions \emph{as irradiated}, and after the annealing steps $T_{ann} = 80^{\,\circ }$, $180^{\,\circ }$ and 280$^{\,\circ }$C.
 The point defects of the initial cluster, which are separated by the distance $d_0$, are generated along the $z$\,direction.
 As expected from figure\,\ref{fig:SigTann}, hardly any diffusion is observed at $T_{ann} = 80^{\,\circ }$C, the width increases by about a factor 3 at 180$^{\,\circ }$C, and the clusters are essentially dissociated at 280$^{\,\circ }$C.
 We note that for $N_t = 2.5\times 10^{12}$\,cm$^{-3}$, shown in figure\,\ref{fig:NtTann}, and 15 point defects in each cluster, the average distance between clusters is $\approx 2\,\upmu $m.

 We now summarise the results of the two models used to describe the annealing measurements:
 Both models provide a description of the dependence $\Delta E_a(T_{ann}) $ for the V$_2$/E205a and the V$_3$/E148\,clusters.
 Whereas the \emph{first order decay model} is a purely phenomenological parametrisation, the \emph{diffusion model} is physics based and gives, within the assumptions made to determine the $\Delta E_a$ values and the limitation of this simple model, insight into the cluster annealing.
 The diffusion-activation energy, $E_D$, is quite different for the two cluster defects: $\approx 0.4 $\,eV for V$_2$/E205a and $\approx 0.15 $\,eV for V$_3$/E148.
 At an annealing temperature of 80$^{\,\circ}$C the diffusion parameter, $D$, for V$_2$/E205a is about a factor 50 smaller than for V$_3$/E148.
 However at 280$^{\,\circ}$C, the highest annealing temperature investigated, they are quite similar.
 As a consequence, the diffusion of the point defects in the V$_3$/E148\,cluster becomes significant only for temperatures above 150$^{\,\circ}$C, whereas for the V$_2$/E205a\,cluster it is already significant at room temperature.
 At the highest annealing temperatures the spread of the points in the cluster approaches the average distance between the clusters.

\section{Conclusions}
 In this paper a new method of analyzing TSC (Thermally Stimulated Current) spectra of radiation-damaged silicon diodes is presented.
 It is based on the Shockley-Read-Hall statistics and can be used to describe both point and cluster defects.
 An ionisation energy, $E_a$, which depends on $f_t$, the fraction of filled point defects in the cluster, is introduced.
 Simple electrostatic model calculations show that a linear dependence $E_a (f_t) = E_0 - \Delta E_a \cdot f_t$ appears to be a reasonable approximation, where $E_0$ is the ionisation energy of the point defect which makes up the cluster, and $\Delta E_a$ the difference in ionisation energy between $f_t \rightarrow 0$ and $f_t = 1$.
 The method is applied to the analysis of TSC\,spectra from n-type silicon diodes irradiated by electrons in the energy range between 3.5 and 27~MeV.
 For the filling of the traps the the p$^+$ implants of the diodes are exposed to green light, so that only acceptor traps contribute to the TSC\,current.

 As a first step the VO$_i$ defect (an electron trap with $E_0 \approx 0.17$\,eV, which is known to be a point defect) is analysed for the diode irradiated by 27~MeV electrons to a fluence of $4.3 \times 10^{14}$\,cm$^{-2}$.
 The value found for $\Delta E_a$ is compatible with zero.
 This confirms that VO$_i$ is a point defect, and also demonstrates the validity of the method for point defects.

 Next, the radiation-induced deep acceptor states with energies between 0.35 and 0.46~eV from the conduction band are investigated.
 They are responsible for part of the generation current, for charge trapping and for changes of the effective space-charge density with irradiation.
 In the analysis it is assumed that four defects, E148, E205a, V$_2$, and V$_3$ contribute to the TSC\,current
 in this energy region.
 For the fits to the TSC spectra, the ionisation energies from the literature and common $\Delta E_a$\,values for V$_2$ and E205a as well as for V$_3$ and E148 had to be assumed.
 Without these or similar not too well justified assumptions, the fits do not converge to stable values.

 To study the dependence of the cluster formation on electron energy, the data for $E_e = 3.5$, 6, 15, and 27\,MeV after annealing for 30~min at 80$^{\,\circ}$C are analysed.
 For the lowest electron energy $\Delta E_a = 0$, whereas for the higher electron energies positive $\Delta E_a$\,values are found.
 After a rapid increase with electron energy, a tendency towards a saturation of $\Delta E_a$ is observed.
 This is taken as evidence for cluster formation above a threshold, which is somewhere between $E_e =3.5 - 6$\,MeV.

 To study the annealing of the clusters, the data for the irradiation with 15\,MeV electrons and the conditions \emph{as irradiated} and after isochronal annealing for 30\,min between $T_{ann} = 80 - 280^{\,\circ}$C are analysed with the assumptions discussed above.
 For both V$_2$/E205 and V$_3$/E148, $\Delta E_a$ is found to decrease with $T_{ann}$, however the decrease for V$_3$/E148 is significantly faster than for V$_2$/E205.
 Two models, a \emph{first order decay model} and a \emph{diffusion model}, provide adequate descriptions of the data.
 Using the latter model the $T_{ann} $ dependence of the diffusion parameters and of the spatial spread of the point defects in the cluster is estimated.
 It is found, that at $T_{ann} = 80^{\,\circ}$C the diffusion constant for V$_2$/E205a is about a factor 50 smaller than for V$_3$/E148, whereas at $T_{ann} = 280^{\,\circ}$C, they are quite similar.
 For the diffusion-activation energy, $E_D$, a value of 0.42\,eV is determined for the V$_2$/E205\,cluster, and  0.15\,eV for the V$_3$/148\,cluster.
 For the highest $T_{ann}$\,values, the spread of the point defects in the clusters approaches the average distance between the clusters.

\section{Acknowledgements}
 This work has been performed in the framework of the CERN-RD50 collaboration and PNII-ID-PCE-2011-3 Nr.\ 72/5.10.2011, and partially funded by CiS, the German BMBF and the Helmholtz Alliance Physics at the Terascale.
 I.~Pintilie and R.~Radu gratefully acknowledge partial funding from the Romanian core program PN18-11 (funded by ANCSI).
 R.~Radu was also partially supported by the Marie Curie Initial Training Network MC-PAD and the fellowship A127881 of the German Academic Exchange Service (DAAD).
 We also thank the teams at the irradiation facilities at
 the Belarusian State University at Minsk, the PTB at Braunschweig, and the ELBE accelerator at Dresden-Rossendorf, where the irradiations have been performed.
 We are grateful to P.~Buhmann and M.~Matysek for maintaining the measurement infrastructure of the Hamburg Detector Laboratory, where the measurements were performed, in an excellent shape.


\end{document}